# High Field (up to 140 kOe) Angle Dependent Magneto Transport of $Bi_2Te_3$ Single Crystals

Rabia Sultana[1,2], P. K. Maheshwari[1,2], Brajesh Tiwari[3] and V.P.S. Awana[1,*]

[1] *CSIR-National Physical Laboratory, New Delhi-110012, India*

[2] *Academy of Scientific and Innovative Research (AcSIR), NPL, New Delhi-110012, India*

[3] *Department of Physics, Institute of Infrastructure Technology Research And Management, Ahmedabad, 380026, India*

## Abstract

We report the angle dependent high field (up to 140 kOe) magneto transport of $Bi_2Te_3$ single crystals, a well-known topological insulator. The crystals were grown from melt of constituent elements via solid state reaction route by self-flux method. Details of crystal growth along with their brief characterization up to 5 Tesla applied field was reported by some of us recently [*J. Magn. Mag. Mater. 428, 213 (2017)*]. The angle dependence of the magneto-resistance (MR) of $Bi_2Te_3$ follows the cos ($\theta$) function i.e., MR is responsive, when the applied field is perpendicular (tilt angle $\theta = 0^o$ and/or $180^o$) to the transport current. The low field ($\pm 10$ kOe) MR showed the signatures of weak anti localization (WAL) character with typical v-type cusp near origin at 5 K. Further, the MR is linear right up to highest applied field of 140 kOe. The large positive MR are observed up to high temperatures and are above 250% and 150% at 140 kOe in perpendicular fields at 50 K and 100 K respectively. Heat capacity $C_P(T)$ measurements revealed the value of Debye temperature ($\theta_D$) to be 135 K. ARPES (angle resolved photoemission spectroscopy) data clearly showed that the bulk $Bi_2Te_3$ single crystal consists of a single Dirac cone.

[*]**Corresponding Author**
Dr. V. P. S. Awana:  E-mail: awana@nplindia.org: Ph. +91-11-45609357, Fax-+91-11-45609310; Homepage: awanavps.webs.com

**Introduction**

Topological insulators (TI's) having dual character of being insulating within their interior bulk due to intrinsic electronic band gap of around 0.3 eV, but conducting at surface via Dirac cone forming surface states are attracting huge attention of the condensed matter physicists including both experimentalists and theoreticians alike [1-8]. One of the most popular and widely studied properties of TI's is large (~ 600%, 2 K, 140 kOe) and non saturating positive magneto resistance (MR) [9-14]. Further, though the MR magnitude is maximum at low temperatures, the same is seen right up to room temperature with non-saturating characteristic and reasonable magnitude [9,13,14]. Another, interesting feature of the observed non saturating linear MR is that the same is highly anisotropic, being maximum when applied field is perpendicular and minimum in parallel [10-17]. Principally, the surface Dirac states are spin-momentum locked and their character depends upon the direction of applied magnetic field, giving rise to anisotropic character [1-8].

As mentioned above, the unusual MR of TI's is highly anisotropic; hence one desires the single crystalline material. The most common examples of topological insulators yet widely studied over last couple of years are $Bi_2Se_3$ and $Bi_2Te_3$ [1, 2] for which single crystals are grown from their high temperature melt ($950^0$C) by slow cooling ($2^0$C/hour) through solid state reaction in a protected environment [11,18]. In current letter, we extend our earlier work [18], whereby we reported crystal growth and magneto transport (up to 50 kOe) of $Bi_2Te_3$. We report here the high field (up to 140 kOe) magneto transport of as grown different batch $Bi_2Te_3$ crystal along with the angle dependence. It is found that the MR is linear and non-saturating up to the applied field of 140 kOe and is highly anisotropic as well. Also, the low field (± 10 kOe) MR is reminiscent of Weyl character with typical ν-type cusp near origin at 5 K. Further, the anisotropic nature of MR, along with brief heat capacity and the angle resolved photoemission spectroscopy (ARPES) data of the studied crystal are reported in addition.

**Experimental Details**

The $Bi_2Te_3$ single crystal used in present study is grown from high temperature ($950^0$C) melt of constituent elements by slow cooling (~$2^0$ C/hour) is protected (sealed in evacuated quartz tubes). Details of crystal growth are given in ref. 18. The crystal structure is determined from room temperature X-ray diffraction (XRD) for powder and single crystalline sample of $Bi_2Te_3$. Magneto transport measurements were carried out on a Quantum Design (QD) Physical Property Measurement System (PPMS) down to 5 K in applied magnetic field of up to 14 Tesla equipped with sample rotator.

**Results and Discussion**

The primary characterization technique used to determine the crystallographic structure was the room temperature X-ray diffraction (XRD). Figure 1 shows the single crystal XRD pattern of the synthesized $Bi_2Te_3$ single crystal. The sharp X-ray diffraction peaks of (0, 0, 3$n$) indicates the growth of the synthesized bulk $Bi_2Te_3$ single crystal along the [001] plane and confirms the high crystalline nature. The synthesized sample is crystallized in rhombohedral structure having rietveld refined lattice parameters as $a = b = 4.3866$ (2) Å and $c = 30.4978$ (13) Å with $R\bar{3}m$ (D5) space group [18]. This phase pure single crystal of $Bi_2Te_3$ is used for further magneto-transport, heat capacity analysis and ARPES (angle resolved photoemission spectroscopy) studies.

Figure 2 shows the resistivity versus tilt angle plot of the studied $Bi_2Te_3$ crystal measured at 5 K under an applied magnetic field of 10 kOe. Here, MR is responsive when the applied field is perpendicular (tilt angle θ = 0° and/or 180°) to the transport current, whereas it remains absent in the parallel field. The peaks and dips are found at about θ = 0°, 180° (perpendicular field component) and θ = 90°, 270° (parallel field component) respectively. Also, the ρ(θ) plot clearly follows the cos(θ) function and shows wide peaks around the perpendicular field component suggesting the existence of quasi-2D Fermi surface in $Bi_2Te_3$ [19]. Clearly, we can see that the resistivity increases with perpendicular component of applied magnetic field whereas, it decreases when the field is changed to parallel component [10-15]. The inset shows the schematic diagram of the sample, the direction of the magnetic field and the transport current.

Figure 3 shows the electrical resistivity versus temperature plot of $Bi_2Te_3$ single crystal measured at different applied magnetic fields. The temperature range varies from 5 K to 50 K whereas, the applied magnetic field ranges from 0 kOe to 100 kOe. Apparently, the temperature dependent electrical resistivity plot for studied $Bi_2Te_3$ single crystal depicts metallic behaviour both in the presence and absence of applied magnetic field. The value of the resistivity increases with increase in temperature from 5 K to 50 K, representing positive temperature coefficient property. Moreover, one can clearly observe that as the applied magnetic field is increased from 0 kOe to 100 kOe, the resistivity value increases gradually. The obtained values of resistivity measured at 50 K under 0 kOe is found to be 0.34 mΩ – cm, whereas the same increases to 0.82 mΩ-cm under an applied field of 100 kOe. Thus, the nature of the plot is in agreement to that as observed in our earlier reported literature [18].

Figure 4 shows the percentage change of magneto–resistance versus applied magnetic field (both parallel and perpendicular) plot of the studied $Bi_2Te_3$ crystal at different temperatures of 5, 50 and 100 K respectively. The MR (%) was calculated using the formula MR = [ρ(H) - ρ(0)] / ρ(0) where, ρ(H) represents the resistivity in the presence of magnetic field and ρ(0) represents the resistivity at zero magnetic field. We observed a large, asymmetric, non-saturating linear positive MR value reaching up to 450% at 5 K under applied perpendicular field of about 140 kOe, whereas it decreases to about 100% when the field is changed from perpendicular to parallel. However, this linear MR(%) is suppressed gradually with increase in temperature from 5 K to 100 K, which is a characteristic of the WAL effect [20]. Thus, the MR (%) value reduces from 440% to 150% as the temperature is increased from 5 K to 100 K. Moreover, as the field is changed from perpendicular to parallel the MR value reduces to 82% at 5 K under an applied magnetic field of 140 kOe. Further, the same reduces from 82% to 43% as the temperature is increased from 5 K to 100 K. Thus, we get a negligible value of MR in parallel component in comparison to the perpendicular component.Consequently, we can say that the MR is angular dependent [10,11,15].

Figure 5 represents the field derivative MR as a function of magnetic field. It can be apparently seen that at lower magnetic field, the MR exhibits a quadratic field response demonstrating a semi classical $H^2$ dependent MR. The quadratic field response at low field is obtained using the equation MR= $A_2H^2$ [19,21-22]. However, as the magnetic field increases the MR exhibits a change from quadratic to linear response. This occurs when the magnetic field is above a characteristic field H* and the dMR/dH saturates to a much reduced slope and finally becomes a constant. Here, the characteristic field H* (Figure 5) is defined as the junction of the fitting lines in lower and higher magnetic field respectively. The linear response of the MR is calculated using the relation MR= $A_1H+ O(H^2)$ [19,21-22]. Thus, the transition of MR from quadratic form at lower field to linear form at higher fields, indicates that the MR is dominated by linear field dependence along with a small quadratic response.

Figure 6 shows the magneto resistance [MR (%)] as a function of applied magnetic field [H(kOe)] for the studied $Bi_2Te_3$ crystal at 5 K. The magnetic field is varied from -10 kOe to 10 kOe range and corresponding magnetoresistance (in units of %) is estimated. Here, as the magnetic field is increased the nature of the curve of positive MR at 5 K tends to become more linear suggesting the presence of Dirac fermions [20]. At lower magnetic field, a typical v-type cusp (sharp MR dip) near origin is clearly observed at 5 K (Figure 6) which gradually tends to broaden with increase in temperature and then finally disappears (Figure 4). Thus, at lower magnetic field a sharp dip-like positive MR is observed in the absence of any magnetic scattering. This type of behaviour near the origin of the graph with a V shape clearly indicates the signatures of weak anti localization (WAL) effect [20, 23-24]. The signature of WAL is claimed to be arising by suppression of back scattering of carriers by the π-Berry phase of topological surface states and helical spin-momentum locking as reported earlier [20-23, 24]. The inset of Figure 6 represents the derivative plot of MR (%) as a function of magnetic field i.e. $dMR(\%)/dH(Oe^{-1})$ vs H(kOe) for the studied $Bi_2Te_3$ crystal at 5 K. Almost constant slope (dMR/dH) as a function of magnetic field also suggests the linear magnetoresistance at high fields. Further, the inset clearly shows a kink at lower magnetic field (~5 kOe) indicating the WAL point.

Figure 7 shows the temperature dependence of the specific heat of the synthesized $Bi_2Te_3$ crystal in temperature ranging from 5 K to 250 K. Red colour line shows the Cp versus T plot fitted using the equation $Cp = \gamma T + \beta T^3 + \delta T^5$, where $\gamma$ represents the Somerfeld coefficient and $\beta$ and $\delta$ represent the coefficients of the phononic contributions. The inset of Figure 7 shows the linear fitted plot of the measured Cp/T versus $T^2$ of the studied $Bi_2Te_3$ single crystal. The Debye temperature ($\theta_D$) calculated using the formula $\theta_D = [234zR/\beta]^{1/3}$ turns out to be 135 K, where z is the number of atoms per formula unit and R is the gas constant. The Cp plot as well as the $\theta_D$ value are in agreement to the earlier reported literature of TI [11, 25-26].

Figure 8 represents the surface state of $Bi_2Te_3$ investigated crystal, using angle resolved photoemission spectroscopy (ARPES). The ARPES data clearly show that the as synthesized bulk $Bi_2Te_3$ single crystal exhibit 3D nature, whose surface state consists of a single Dirac cone at the Γ point. Accordingly, the ARPES data matches well with previously reported literature [1-8, 27]. The energy distribution map (EDM) is measured at the zone center from $Bi_2Te_3$ at a sample temperature of 20 K using 70 eV photon energy and *p*-polarized light. Dirac point is located at a binding energy of 300 meV below the Fermi level as shown by the green arrow on the EDM. The data are taken at Elettra synchrotron light source.

**Conclusion**

In conclusion, high quality reproducible single crystals of $Bi_2Te_3$ are synthesized. High field angle dependent magneto transport behaviour is observed in the synthesized $Bi_2Te_3$ single crystal. The low field MR showed the WAL effect whereas, the high field showed linear positive MR reaching up to 450% along with anisotropc nature at 5 K. The MR turns from quadratic (at lower field) to linear suggesting the existence of Dirac fermions. Heat capacity $C_P(T)$ measurements revealed the value of Debye temperature ($\theta_D$) to be 135 K. The ARPES clearly showed the presence of a single Dirac cone at the $\Gamma$ point of the studied $Bi_2Te_3$ single crystal.

**Acknowledgements**

The authors from CSIR-NPL would like to thank their Director NPL, India, for his keen interest in the present work. This work is financially supported by DAE-SRC Outstanding Investigator Award Scheme on Search for New Superconductors. Rabia Sultana thanks CSIR, India, for research fellowship and AcSIR - NPL for Ph.D. registration. The authors thank S. Thirupathaiah in the department of SSCU, IISc for the ARPES measurement.

**Figure Captions**

**Figure 1:** X-ray diffraction pattern for $Bi_2Te_3$ single crystal.

**Figure 2:** Angular dependent resistivity plot for the studied $Bi_2Te_3$ single crystal measured at 5 K and 10 kOe magnetic field. Inset shows the schematic diagram of the sample, the direction of the magnetic field and the transport current.

**Figure 3:** Temperature dependent electrical resistivity plot of $Bi_2Te_3$ single crystal at different applied magnetic fields.

**Figure 4:** MR (%) as a function of both perpendicular as well as parallel magnetic field (H) for the studied $Bi_2Te_3$ crystal at different temperatures.

**Figure 5:** Field derivative of MR (%) as a function of perpendicular magnetic field (H) for $Bi_2Te_3$ at different temperatures.

**Figure 6:** Percentage change of magneto-resistance as a function of applied magnetic field showing WAL signatures of $Bi_2Te_3$ single crystal at 5 K. Inset shows the derivative of WAL signature of $Bi_2Te_3$ single crystal at 5 K.

**Figure 7:** Heat capacity (Cp) versus temperature (T) plot for $Bi_2Te_3$ single crystal. Inset shows the Cp/T versus $T^2$ plot of the studied $Bi_2Te_3$ crystal.

**Figure 8:** ARPES data of $Bi_2Te_3$ showing the Dirac point.

Fig. 1

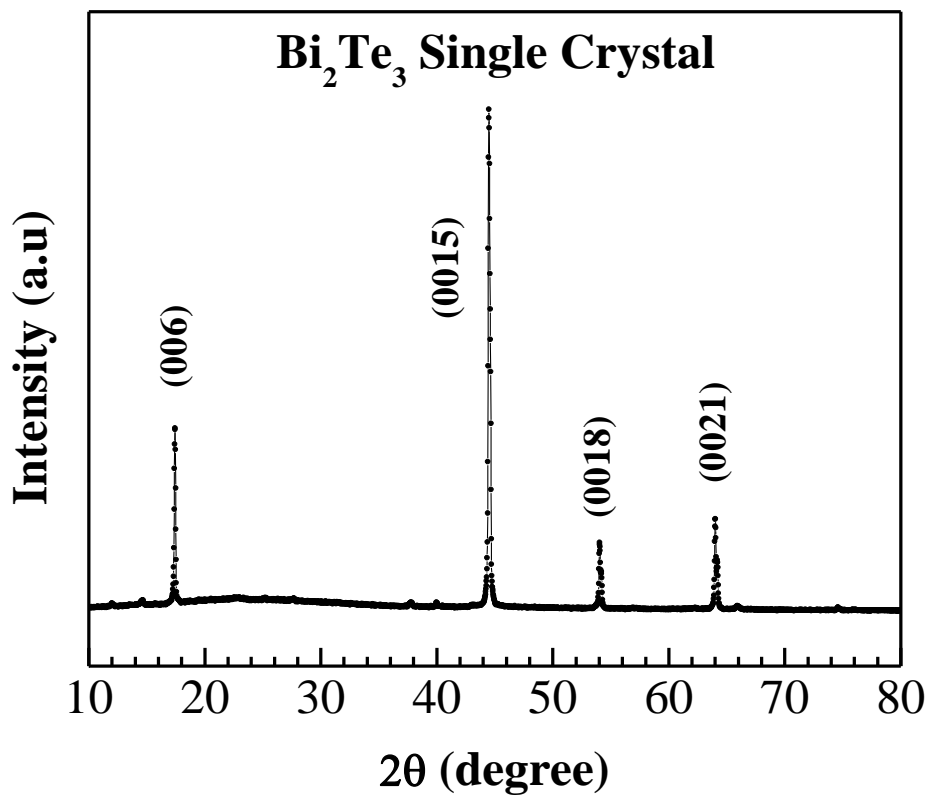

Fig. 2

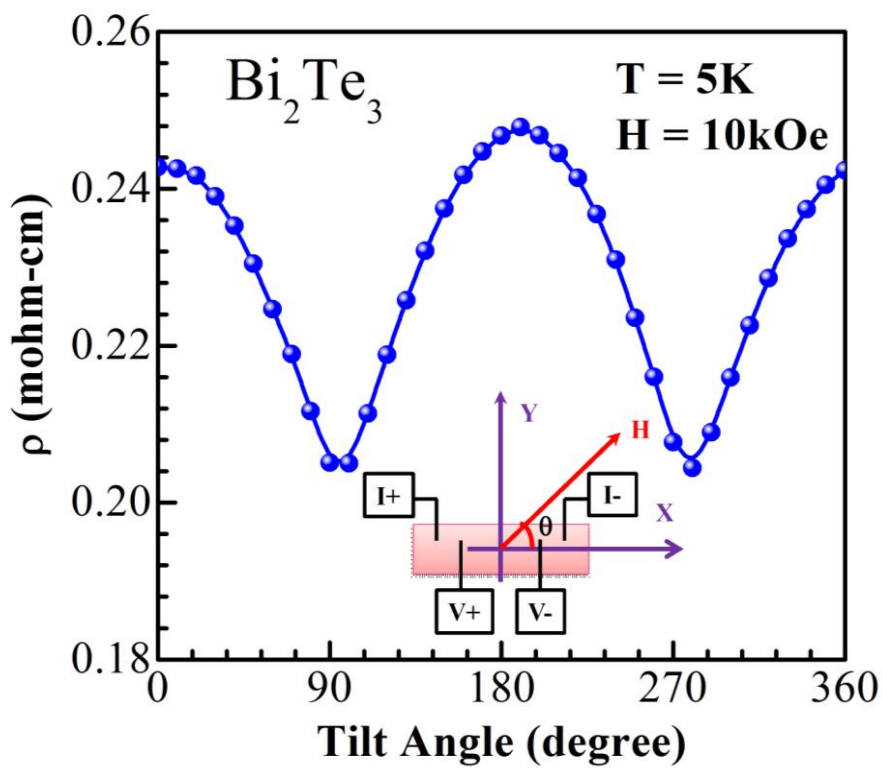

Fig. 3

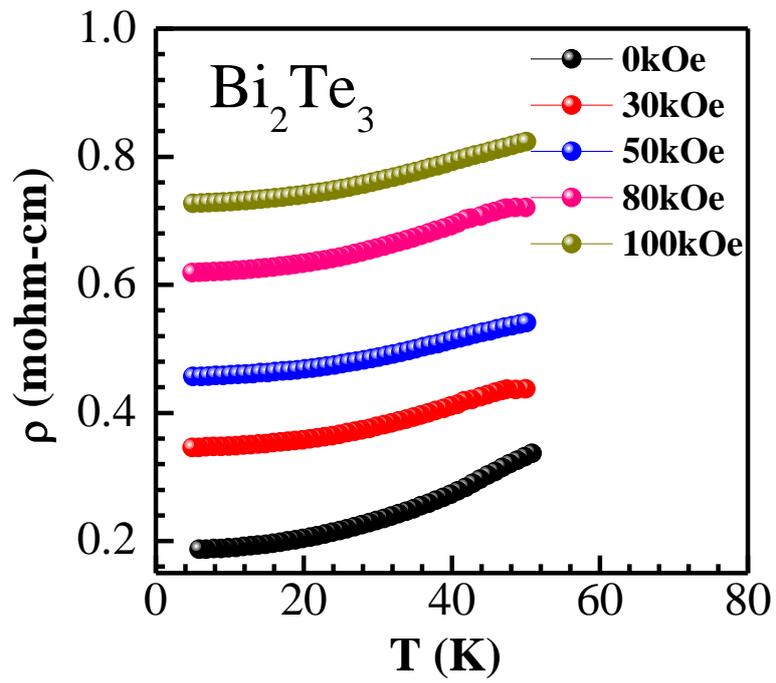

Fig. 4

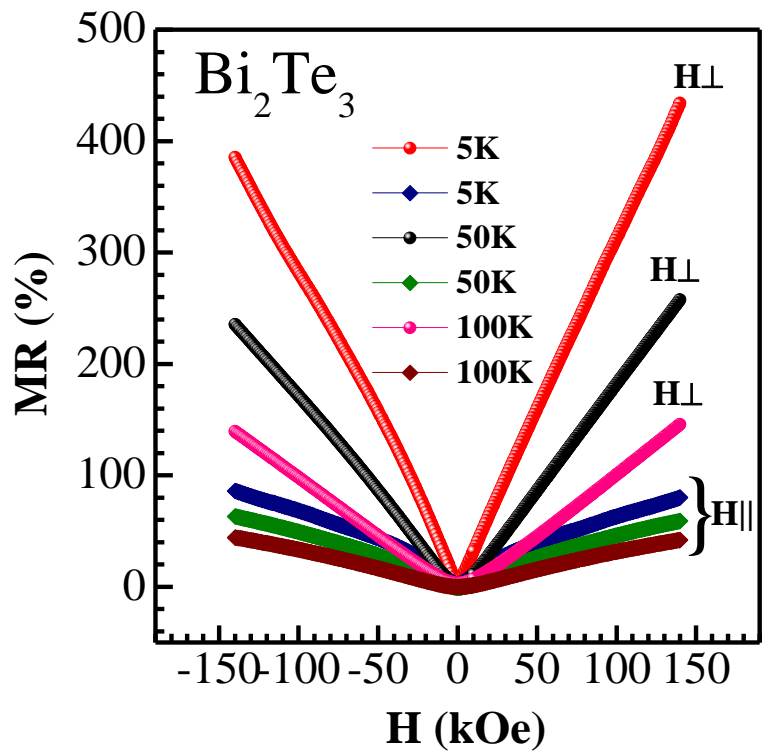

Fig. 5

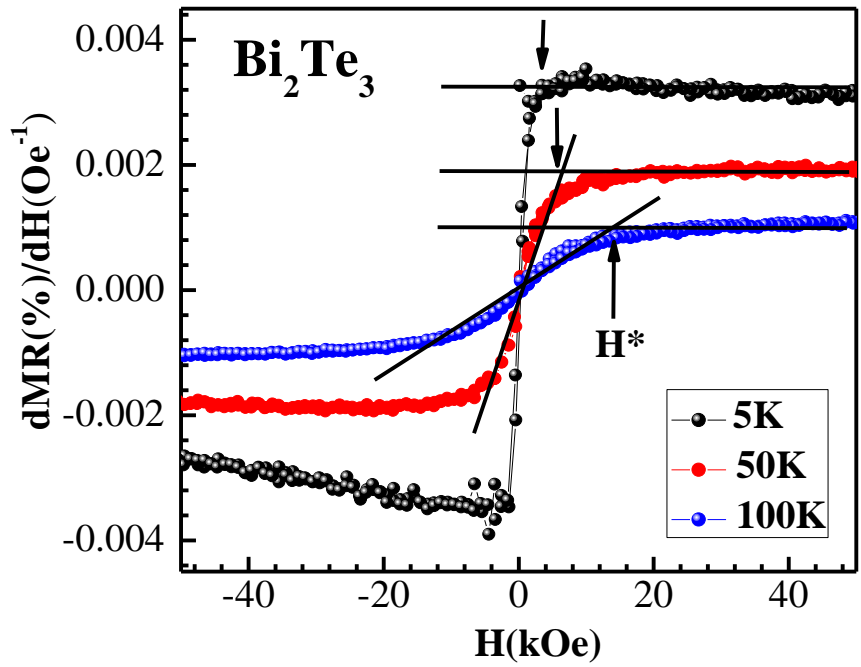

Fig. 6

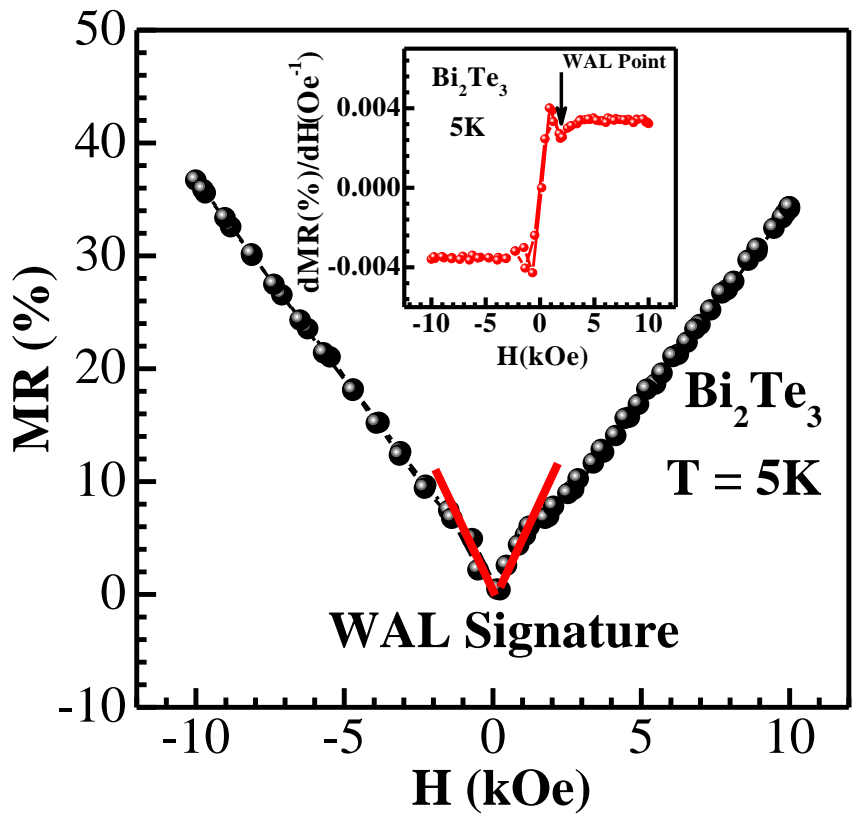

Fig.7

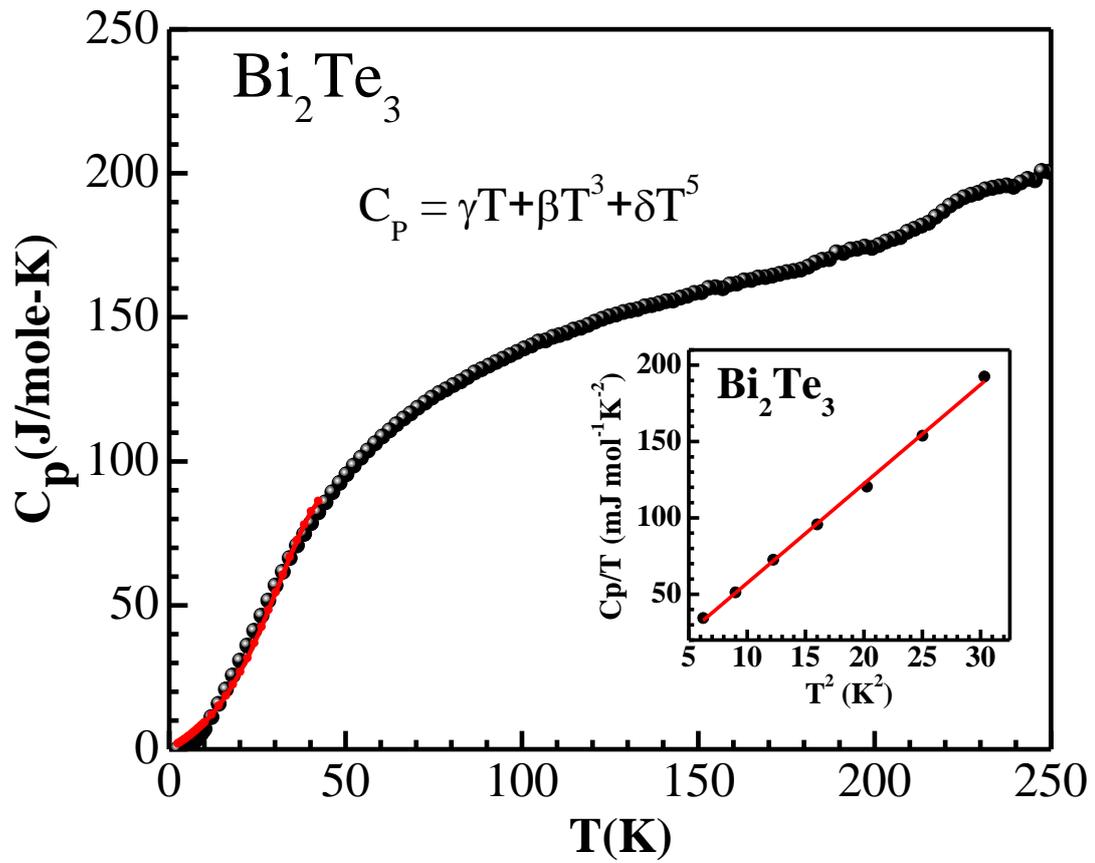

Fig. 8

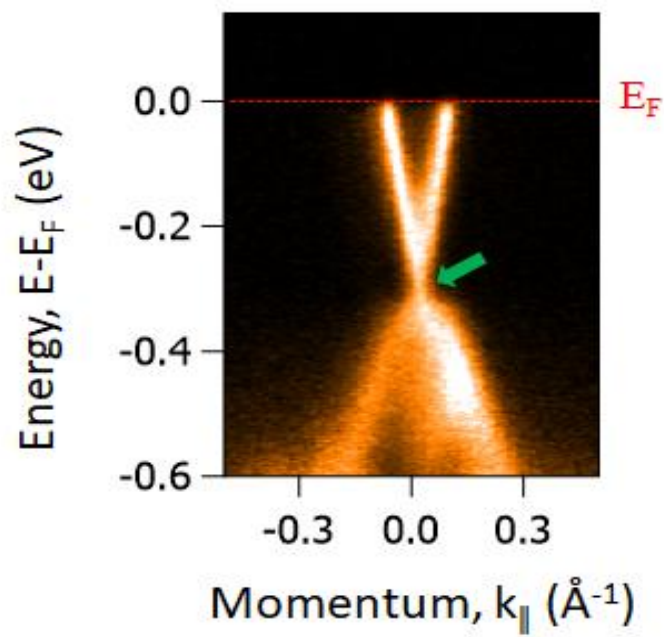